\documentclass[nohyper]{JHEP3}
\usepackage{enumerate,amssymb}
\renewcommand{\a}{\alpha}
\renewcommand{\b}{\beta}
\newcommand{\g}{\gamma}           
\renewcommand{\d}{\delta}         
\newcommand{\e}{\varepsilon}

\newcommand{\ki}{\chi}
\newcommand{\la}{\lambda}

\newcommand{\p}{\psi}             

\newcommand{\s}{\sigma}           \renewcommand{\S}{\Sigma}
         
\newcommand{\f}{{\phi}}           \newcommand{\F}{{\Phi}}

\newcommand{\eps}{{\epsilon}}

\newcommand{\ca}{{\cal A}}
\newcommand{\cb}{{\cal B}}
\newcommand{\cc}{{\cal C}}
\newcommand{\cd}{{\cal D}}

\newcommand{\cf}{{\cal F}}
\newcommand{\cg}{{\cal G}}

\newcommand{\cR}{{\cal R}}

\newcommand{\cw}{{\cal W}}

\newcommand{\be}{\begin{equation}}
\newcommand{\ee}{\end{equation}}
\newcommand{\eqn}[1]{\label{#1}\end{equation}}
\newcommand{\equ}[1]{(\ref{#1})}

\newcommand{\bea}{\begin{eqnarray}}
\newcommand{\eea}{\end{eqnarray}}
\newcommand{\eqan}[1]{\label{#1}\end{eqnarray}}

\newcommand{\ba}{\begin{array}}
\newcommand{\ea}{\end{array}}

\newcommand{\nn}{\nonumber}

\newcommand{\loco}{{\mathop{ \, \rule[-.06in]{.2mm}{3.8mm}\,}}}
\newcommand{\doubar}{{{\loco}\!{\loco}}}
\newcommand{\az}{{\bf{z}}}
\newcommand{\au}{{\bf{u}}}
\newcommand{\av}{{\bf{v}}}
\newcommand{\aw}{{\bf{w}}}

\newcommand{\da}{{\dot{\alpha}}}
\newcommand{\db}{{\dot{\beta}}}
\newcommand{\dg}{{\dot{\gamma}}}
\newcommand{\dd}{{\dot{\delta}}}

\newcommand{\ta}{{\mbox{\tiny{A}}}}
\newcommand{\tb}{{\mbox{\tiny{B}}}}
\newcommand{\tc}{{\mbox{\tiny{C}}}}
\newcommand{\td}{{\mbox{\tiny{D}}}}
\newcommand{\te}{{\mbox{\tiny{E}}}}
\newcommand{\tf}{{\mbox{\tiny{F}}}}
\newcommand{\tg}{{\mbox{\tiny{G}}}}

\renewcommand{\gg}{\mathfrak{g}}
\newcommand{\hh}{{\mathfrak{h}}}
\renewcommand{\tt}{{\mathfrak{t}}}

\newcommand{\vd}{\textrm{d}}
\newcommand{\bP}{\bar{P}}
\newcommand{\bs}{\bar{\s}}
\newcommand{\bp}{\bar{\p}}
\newcommand{\bla}{\bar{\la}}

\title{Equations of motion for $N=4$ supergravity with antisymmetric
tensor from its geometric description in central charge superspace}
\author{S\'ebastien GURRIERI and Annam\'aria KISS\\
Centre de Physique Th\'eorique, CNRS Luminy, Case 907\\
F-13288 Marseille -- Cedex 9\\
E-mail: gurrieri@cpt.univ-mrs.fr, kiss@cpt.univ-mrs.fr
}
\abstract{We consider the geometrical formulation in central charge
superspace of the $N=4$ supergravity containing
an antisymmetric tensor gauge field. The theory is on-shell, so
clearly, the constraints used for the
identification of the multiplet together with the superspace Bianchi
identities imply equations of motion
for the component fields. We deduce these equations of motion in
terms of supercovariant quantities and
then, we give them in terms of component fields. These equations of
motion, deduced from the geometry,
without supposing the existence of a Lagrangian, are found to be the
same as those derived from the
Lagrangian given in the component formulation of this $N=4$
supergravity multiplet by Nicolai and
Townsend.}
\preprint{JHEP 0202 : 040, 2002}
\begin{document}
%%%%%%%%%%%%%%%%%%%%%%%%%%%%%%%%%%%%%%%%%%%%%%%%%%%
\section{Introduction}
%%%%%%%%%%%%%%%%%%%%%%%%%%%%%%%%%%%%%%%%%%%%%%%%%%%

The $N=4$ supergravity theory containing an antisymmetric tensor was given by
Nicolai and Townsend \cite{NT81} already in the early eighties. The
superspace formulation of the corresponding multiplet,  which we call
the N-T multiplet in the following, encountered a number of problems
identified in \cite{Gat83} and overcome in \cite{GD89} introducing
external Chern-Simons forms for the graviphotons. Recently, a concise
geometric formulation was given for this supergravity theory in
central charge superspace \cite{GHK01}.

The geometric approach adopted and described in detail in
\cite{GHK01} was based on the
superspace soldering mechanism involving gravity and 2--form geometries in
central charge superspace \cite{AGHH99}. This soldering procedure
allowed to identify
various gauge component fields of the one and the same multiplet in two
distinct geometric structures: graviton, gravitini and graviphotons in the
gravity sector and the antisymmetric tensor in the 2--form sector.
Supersymmetry and central charge transformations of the component fields were
deduced using the fact that in the geometric approach these transformations
are identified on the same footing with general space-time coordinate
transformations as superspace diffeomorphisms on the central charge
superspace. Moreover, the presence of graviphoton Chern-Simons forms in the
theory was interpreted as an intrinsic property of central charge superspace
and a consequence of the superspace soldering mechanism.

The aim of the present paper is to emphasize that the geometric
description in \cite{GHK01} is on-shell,
that is the constraints used to identify the component fields of the
N-T supergravity multiplet imply
also the equations of motion for these fields. Therefore, we begin
with recalling the essential points in
the identification of component fields \cite{GHK01} specifying the
constraints we use. Then, in section
3, we derive the equations of motion directly from constraints and
Bianchi identities, without any
knowledge about a Lagrangian. Finally, we compare these equations of
motion with those found from the
component Lagrangian given in the original article by Nicolai and
Townsend \cite{NT81}.

%%%%%%%%%%%%%%%%%%%%%%%%%%%%%%%%%%%%%%%%%%%%%%%%%%%%%%%%%%%%%
\section{Constraints and identification of component fields}
%%%%%%%%%%%%%%%%%%%%%%%%%%%%%%%%%%%%%%%%%%%%%%%%%%%%%%%%%%%%%

In this section we recall the essential results of \cite{GHK01}
concerning the identification of the
components of the N-T multiplet. Conventions, notations and general
ideas about geometrical description of
supergravity theories, central charge superspace and soldering
mechanism are detailed in \cite{GHK01} and
in references mentioned there.

Recall that in geometrical formulation of supergravity theories
the basic dynamic variables are chosen to be the vielbein and the
connection. Considering central charge superspace this framework
provides a unified geometric identification of graviton, gravitini
and graviphotons in the frame $E^\ca=(E^a,E_\ta^\a,E^\ta_\da,
E^\au)$, where $a$, $\a$, $\da$ denote the usual vector and Weyl
spinor indices, while capital indices $\ta$ count the number of
supercharges and boldface indices $\au = 1..6$ the number of
central charges: \be E^a\doubar\ =\ \vd x^m e_m{}^a~,\quad
E_\ta^\a\doubar\ =\ \frac{1}{2}\,\vd x^m \p_m{}_\ta^\a~, \quad
E^\ta_\da\doubar\ =\ \frac{1}{2}\,\vd x^m \bar{\p}_m{}^\ta_\da~,
\quad E^\au\doubar\ =\ \vd x^m v_m{}^\au~, \eqn{frame} while the
antisymmetric tensor can be identified in a superspace 2--form
$B$: \be B\doubar\ =\ \frac{1}{2}\,\vd x^m \vd x^n b_{nm}. \eqn{B}
The remaining component fields, a real scalar and 4 helicity 1/2
fields, are identified in the supersymmetry transforms of the
vielbein and 2--form, that is in torsion ($T^\ca=DE^\ca$) and
3--form ($H=dB$) components. The Bianchi identities satisfied by
these objects are \be DT^\ca\ =\ E^\cb
R_\cb{}^\ca\,,\qquad\qquad\vd H\ =\ 0\,, \ee or displaying the
3--form and 4--form coefficients \bea
\left(_{\cd\cc\cb}{}^\ca\right)_T\quad&:&\quad E^\cb E^\cc E^\cd
\left(\cd_\cd T_{\cc\cb}{}^\ca+T_{\cd\cc}{}^\cf
T_{\cf\cb}{}^\ca-R_{\cd\cc\cb}{}^\ca
\right)=0,\\
\left(_{\cd\cc\cb\ca}\right)_H\quad&:&\quad
E^\ca E^\cb E^\cc E^\cd \left(2\cd_\cd
H_{\cc\cb\ca}+3T_{\cd\cc}{}^\cf H_{\cf\cb\ca}\right)=0.
\eea

By putting constraints on torsion and 3--form we have to solve two problems at
the same time: first, we have to reduce the huge number of superfluous
independent fields contained in these geometrical objects, and second, we have
to make sure that the antisymmetric tensor takes part of the {\it
same} multiplet
as $e_m{}^a$, $\p_m{}_\ta^\a$, $\bar{\p}_m{}^\ta_\da$, $v_m{}^\au$
(soldering mechanism).

Indeed, the biggest problem in finding a geometrical description of
an off-shell supersymmetric theory is
to find suitable covariant constraints which do reduce this number
but do not imply equations of motion
for the remaining fields. There are several approaches to this
question. One of them is based on
conventional constraints, which resume to suitable redefinitions of
the vielbein and connection and which
do not imply equations of motion \cite{GGMW84b}. However, such
redefinitions leave intact torsion
components with 0 canonical dimension and there is no general recipes
to indicate how these torsion
components have to be constrained. A simplest manner of constraining
0 dimensional torsion components
together with conventional constraints give rise to the so-called
{\it natural constraints}, which were
analyzed in a systematic way both in ordinary extended superspace
\cite{Mul86b} and in central charge
superspace \cite{Kis00}. Another approach is that of the {\it
geometrical constraints} or integrability
conditions \cite{CL86}, \cite{Cha88b}, which in addition to
constraints on the torsion involves also
constraints on the curvature, and which possess many of the
integrability properties as found in the
self-dual Yang-Mills system. There is no equivalence established
between the two approaches,
nevertheless, both imply the same second order conformal type
equations of motion for $N>4$ \cite{GG83},
\cite{CL86}.

The geometrical description of the N-T multiplet is based on a set of
natural constraints in central
charge superspace with structure group $SL(2,\mathbb{C})\otimes
U(4)$. The generalizations of the
canonical dimension 0 ``trivial constraints" \cite{Mul86b} to central
charge superspace are

\be
  T{^\tc_\g}{^\tb_\b}{^a_{}}\ =\ 0~, \qquad
  T{^\tc_\g}{^\db_\tb}{^a}\ =\ -2i\d{^\tc_\tb}(\s{^a}\eps){_\g}{^\db}~, \qquad
  T{^\dg_\tc}{^\db_\tb}{^a}\ =\ 0~,
\eqn{T01}
\be
T{^\tc_\g}{^\tb_\b}{^\au}\ =\ \eps_{\g\b}T^{[\tc\tb]\au}~,\qquad
T^\tc_\g{}^\db_\tb{}^\au\ =\ 0~, \qquad T{^\dg_\tc}{^\db_\tb}{^\au}\ =\
\eps^{\dg\db}T{_{[\tc\tb]}}{}^\au~.
\eqn{T02}

As explained in detail in the article \cite{GHK01}, the soldering is
achieved by requiring some analogous,
``mirror''-constraints for the 2--form sector. Besides the -1/2
dimensional constraints
$H^\tc_\g{}^\tb_\b{}^\ta_\a = H^\tc_\g{}^\tb_\b{}^\da_\ta =
H^\tc_\g{}^\db_\tb{}^\da_\ta =
H^\dg_\tc{}^\db_\tb{}^\da_\ta=0$, we impose
\be
  H{^\tc_\g}{^\tb_\b}{_a}\ =\ 0~, \qquad
  H{^\tc_\g}{^\db_\tb}{_a}\ =\ -2i\d{^\tc_\tb}(\s{_a}\eps){_\g}{^\db}L~, \qquad
  H{^\dg_\tc}{^\db_\tb}{_a}\ =\ 0~,
\eqn{H01}
\be
H{^\tb_\b}{^\ta_\a}_\au=\eps_{\b\a}H_\au{}^{[\tb\ta]} \,,\qquad
H^\tc_\g{}^\db_\tb{}_\au =0\,,\qquad
H^\db_\tb{}^\da_\ta{}_\au=\eps^{\db\da}H_{\au[\tb\ta]}\,,
\eqn{H02}
with $L$ a real superfield. The physical scalar $\f$ of the
multiplet, called also graviscalar, is
identified in this superfield, parameterized as $L=e^{2\f}$. In turn,
the helicity 1/2 fields, called also
gravigini fields, are identified as usual \cite{How82}, \cite{GG83},
\cite{Gat83} in the 1/2--dimensional
torsion component
\be
\eps^{\b\g}T^\tc_\g{}^\tb_\b{}^\ta_\da\ =\ 2T^{[\tc\tb\ta]}{}_\da,\qquad
\eps_{\db\dg}T_\tc^\dg{}_\tb^\db{}_\ta^\a\ =\ 2T_{[\tc\tb\ta]}{}^\a.
\ee

The scalar, the four helicity 1/2 fields, together with the
gauge--fields defined in \equ{frame} and
\equ{B} constitute the N-T on-shell N=4 supergravity multiplet.
However, the 0 dimensional natural
constraints listed above are not sufficient to insure that these are
the {\it only} fields transforming
into each-other by supergravity transformations. The elimination of a
big number of superfluous fields is
achieved by assuming the constraints
\be
\cd^{\td\a} T_{[\tc\tb\ta]\a}\ =\ 0\,,\qquad\qquad
\cd_{\td\da} T^{[\tc\tb\ta]\da}\ =\ 0\,,
\eqn{dt}
and
\be
T_{\az\cb}{}^\ca=0,
\eqn{tz}
as well as all possible compatible conventional
constraints\footnote{see equations \equ{conv} in the
appendix} \cite{Mul86b}, \cite{Kis00}.

It is worthwhile to note that even at this stage the assumptions are
not sufficient to constrain the
geometry to the N-T multiplet. This setup allows to give a
geometrical description at least of the
coupling of $N=4$ supergravity with antisymmetric tensor to six
copies of $N=4$ Yang-Mills \cite{Cha81a}.
Nevertheless, they are strong enough to put the underlying multiplet
on-shell. In order to see this, one
can easily verify that the dimension 1 Bianchi identities
$\left(^\dd_\td{}^\dg_\tc{}^\tb_\b{}^\a_\ta\right)_T$ and
$\left(^\dd_\td{}^\dg_\tc{}_\tb^\db{}_\da^\ta\right)_T$ for the
torsion as well as their complex
conjugates imply
\be
\begin{array}{lcl}
\cd^\td_{\d} T_{[\tc\tb\ta]\a}\ =\ -i
\d_{\tc\tb\ta}^{\td\te\tf}G_{(\d\a)[\te\tf]}\,,&\qquad&
\cd_\td^{\dd} T^{[\tc\tb\ta]\da}\ =\ -i
\d^{\tc\tb\ta}_{\td\te\tf}G^{(\dd\da)[\te\tf]}\,,\\[4mm]
\cd_\td^\dd T_{[\tc\tb\ta]\a}\ =\ P_\a{}^\dd{}_{[\td\tc\tb\ta]}\,,&&
\cd^\td_\d T^{[\tc\tb\ta]\da}\ =\ P_\d{}^\da{}^{[\td\tc\tb\ta]}\,,
\end{array}
\label{dim1}
\ee
with $G$ and $P$ a priori some arbitrary superfields. Let us write
one of the last relations as
\be
\sum_{\td\tc}\cd_\td^\dd T_{[\tc\tb\ta]\a}\ =\ 0,
\ee
take its spinorial derivative $\cd^\te_\e$
\be
\sum_{\td\tc}\left(\left\{\cd^\te_\e,\cd_\td^\dd\right\} T_{[\tc\tb\ta]\a}
-\cd_\td^\dd\left(\cd^\te_\e T_{[\tc\tb\ta]\a}\right)\right)\ =\ 0,
\ee
and observe that the antisymmetric part of this relation in the
indices $\e$ and $\a$ gives rise to Dirac
equation for the helicity 1/2 fields, that is
$\partial^{\a\dd}T_{[\tc\tb\ta]\a}=0$ in the linear
approach.

It turns out, that there is a simple solution of both the Bianchi
identities of the torsion and 3--form,
which satisfies the above mentioned constraints and reproduce the N-T
multiplet. The non-zero torsion and
3--form components for this solution are listed in the appendix, we
will concentrate here on its
properties which are essential for the identification of the
multiplet and the derivation of the
equations of motion for the component fields.

Recall that the particularity of this solution is based on the
identification of the scalar superfield $\f$ in the 0 dimensional
torsion and 3--form components containing a central charge index
\be T^{[\tb\ta]\au}\ =\ 4e^\f\tt^{[\tb\ta]\au}\,, \qquad
T_{[\tb\ta]}{}^\au\ =\ 4e^\f\tt_{[\tb\ta]}{}^\au\,, \ee \be
H_\au{}^{[\tb\ta]}\ =\ 4e^\f\hh_\au{}^{[\tb\ta]}\,, \qquad
H_{\au[\tb\ta]}\ =\ 4e^\f\hh_{\au[\tb\ta]}\,, \ee with
$\tt^{[\tc\tb]\au}$, $\tt_{[\tc\tb]}{}^\au$,
$\hh_\au{}^{[\tb\ta]}$, $\hh_{\au[\tb\ta]}$ constant matrix
elements satisfying the self--duality relations \be
\tt^{[\td\tc]\au} \ =\  \frac{q}{2}\,
\e^{\td\tc\tb\ta}\tt_{[\tb\ta]}{}^\au\,,\qquad\qquad
\hh_\au{}^{[\tb\ta]} \ =\
\frac{q}{2}\,\hh_{\au[\td\tc]}\e^{\td\tc\tb\ta}\qquad\textrm{with}\quad
q=\pm1. \eqn{du0}

Note, that these relations look similar to {\it some} of the
properties of the 6
real, antisymmetric $4\times 4$ matrices $\a^n$, $\b^n$, $(n=1,2,3)$
of $SU(2)\otimes
SU(2)$ %\cite{GSO87},
\cite{CSF78}, \cite{FS78}, which appear in the component
formulation of $N=4$ supergravity theories. Indeed, if we define the
matrices \be
\tt\doteq\left(
\begin{array}{c}\tt^{[\td\tc]}{}^\au\\
\tt_{[\td\tc]}{}^\au\end{array}\right)\,,
\qquad \hh\doteq\left(
\begin{array}{cc}\hh_\au{}_{[\td\tc]}&
\hh_\au{}^{[\td\tc]}\end{array}\right)\qquad\textrm{and}
\ee
\be
\S\doteq\left(
\begin{array}{cc}0&\frac{q}{2}\e^{\td\tc\tb\ta}\\
\frac{q}{2}\e_{\td\tc\tb\ta}&0\end{array}\right)\,,
\qquad \mathbf{1}\doteq\left(
\begin{array}{cc}\frac{1}{2}\d^{\td\tc}_{\tb\ta}&0\\
0&\frac{1}{2}\d_{\td\tc}^{\tb\ta}\end{array}\right)\qquad\textrm{satisfying}
\qquad \S^2=\mathbf{1}\,, \ee then the properties of the matrix
elements $\tt^{[\tc\tb]\au}$, $\tt_{[\tc\tb]}{}^\au$,
$\hh_\au{}^{[\tb\ta]}$, $\hh_{\au[\tb\ta]}$ can be written in a
compact way as follows: \be \S\tt\ =\ \tt\,,\qquad\qquad\hh\S\ =\
\hh\,, \eqn{du} \be \tt\hh\ =\
\mathbf{1}+\S\,,\qquad\qquad(\hh\tt)_\au{}^\av\ =\ 2\d^\av_\au\,.
\eqn{rel_ht}

%Recall, however that we didn't fix a priori the number of the central
%charge coordinates in the
%superspace. The interesting feature of the above properties is that
%taking the trace of relations in
%\equ{rel_ht} one finds $\d^\au_\au=6$, that is the number of central
%charge indices - and thus, the number
%of the vector gauge--fields $v_m{}^\au$ - is determined to be 6.

These matrices serve as converters between the central charge
basis (indices \au) and the $SU(4)$ basis in the antisymmetric
representation (indices $[\td\tc]$). In particular, for the 6
vector gauge fields $v_m{}^\au$ of the N-T multiplet there is an
alternative basis, called the $SU(4)$ basis, defined by \be \left(
\begin{array}{cc}V_m{}_{[\td\tc]}&V_m{}^{[\td\tc]}\end{array}
\right)\ \doteq\
v_m{}^\au\left(
\begin{array}{cc}\hh_\au{}_{[\td\tc]}&\hh_\au{}^{[\td\tc]}\end{array}
\right)\,, \ee where the two components are connected by the
self--duality relations\footnote{Note the similarity between
these self--duality relations and the reality conditions employed
in the description of the $N=4$ Yang-Mills theory \cite{Soh78a}}
\be V_m{}^{[\td\tc]} \ =\  \frac{q}{2}\,
\e^{\td\tc\tb\ta}V_m{}_{[\tb\ta]}. \ee

Moreover, if we look at self--duality properties \equ{du0} as the
lifting and lowering of $SU(4)$ indices
with metric $\frac{q}{2}\e_{\td\tc\tb\ta}$, then a corresponding
metric in the central charge basis can
be defined by
\be
\gg_{\av\au} \ =\
\frac{q}{2}\,\e_{\td\tc\tb\ta}\,\hh_\av{}^{[\td\tc]}\,\hh_\au{}^{[\tb\ta]},
\qquad\quad
\gg^{\av\au}\ =\
\frac{q}{2}\,\e_{\td\tc\tb\ta}\,\tt^{[\td\tc]}{}^\av\,\tt^{[\tb\ta]}{}^\au,
\eqn{metric}
satisfying
\be
\gg_{\au\aw}\gg^{\aw\av} \ = \ \delta_\au^\av~.
\ee
These are the objects which are found to connect torsion and 3--form
components containing at least one
central charge index
\be
H_{\cd\cc\au}\ =\ T_{\cd\cc}{}^\az\gg_{\az\au},\qquad \quad
T_{\cd\cc}{}^\au\ =\ H_{\cd\cc\az}
\gg^{\az\au},
\ee
insuring the soldering of the two geometries.

The four helicity 1/2 fields $T_{[\tc\tb\ta]\a}$,
$T^{[\tc\tb\ta]\da}$ turn out to be equivalent to the
fermionic partner of the graviscalar $\f$
\be
\la^\ta_\a\ = 2\cd^\ta_\a\f\,,\qquad\qquad\bla_\ta^\da\ = 2\cd_\ta^\da\f\,,
\ee
since the following duality relation holds in this $N=4$ case:
\be
T_{[\tc\tb\ta]\a}\ =\ q\e_{\tc\tb\ta\tf}\bla_\a^\tf\,
\qquad\qquad T^{[\tc\tb\ta]\da}\ =\ q\e^{\tc\tb\ta\tf}\la^\da_\tf\,.
\eqn{Tla}

It is the soldering mechanism between the geometry of supergravity
and the geometry of the 2--form, that
determines how the superfields $G$ and $P$ in the spinorial
derivatives of this helicity 1/2 fields
\equ{dim1} are related to the component fields of the multiplet. In
particular, we find that the
superfields $G$ are related to the covariant field strength of the
graviphotons $F_{ba}{}^\au$
\be
G_{(\b\a)}{}_{[\tb\ta]}\ = \
-2ie^{-\f}F_{(\b\a)}{}^\au\hh_\au{}_{[\tb\ta]}\,,\label{G-F}
%\ =\ -2iL^{-1/2}F_{ba}{}_{[\tb\ta]}{}^{(+)}
\qquad\qquad
G^{(\db\da)}{}^{[\tb\ta]}\ =\
-2ie^{-\f}F^{(\db\da)}{}^\au\hh_\au{}^{[\tb\ta]}\,,%\label{G-F2}
%\ =\ -2iL^{-1/2}F_{ba}{}^{[\tb\ta]}{}^{(-)}\, ,
\ee
whereas the superfields $P$ contain the dual field strength of the
antisymmetric tensor and the derivative
of the scalar:
\bea
\cd^\td_\d T^{[\tc\tb\ta]\da}&=&q\e^{\td\tc\tb\ta}P_\d{}^\da\,,\
\textrm{ with }\
P_a\ =\ 2i\cd_a\f+e^{-2\f}H^*_a-\frac{3}{4}\la^\ta\s_a\bla_\ta\,,\label{P}\\
\cd_\td^\da T_{[\tc\tb\ta]\d}&=&q\e_{\td\tc\tb\ta}\bP_\d{}^\da\,,\
\textrm{ with }\
\bP_a\ =\ 2i\cd_a\f-e^{-2\f}H^*_a+\frac{3}{4}\la^\ta\s_a\bla_\ta\,,\label{barP}
\eea
where we can note that the relations
\be
P_a+\bP_a\ =\ 4i\cd_a\f\,,\qquad\qquad P_a-\bP_a\ =\
2e^{-2\f}H^*_a-\frac{3}{2}\la^\ta\s_a\bla_\ta
\eqn{PP}
allow to separate the dual field strength of the antisymmetric tensor
and the derivative of the scalar (as
"real" and "imaginary" part of $P$).

Finally, let us precise that the representation of the structure
group in the central charge sector is
trivial, $\F_\au{}^\az=0$, while the $U(4)$ part $\F^\tb{}_\ta$ of
the $SL(2,\mathbb{C})\otimes U(4)$
connection
\be
\F^\tb_\b{}^\a_\ta\ =\ \d^\tb_\ta\F_\b{}^\a+\d_\b^\a\F^\tb{}_\ta
\qquad\quad\F_\tb^\db{}_\da^\ta\ =\
\d_\tb^\ta\F^\db{}_\da-\d^\db_\da\F^\ta{}_\tb
\ee
is determined to be
\be
\F^\tb{}_\ta\ =\ a^\tb{}_\ta\ +\ \ki^\tb{}_\ta\,,
\ee
with $a^\tb{}_\ta$ pure gauge and $\ki^\tb{}_\ta$ a supercovariant
1--form on the superspace with
components
\bea
\ki_c{}^\tb{}_\ta&=&\frac{1}{4}\d^\tb_\ta\left(ie^{-2\f}H^*_c-\frac{i}{4}\la^\tf\s_c\bla_\tf\right)
-\frac{i}{8}(\la^\tb\s_c\bla_\ta)\,,\nn\\[2mm]
\ki^\tc_\g{}^\tb{}_\ta&=&\frac{1}{4}\d^\tb_\ta\la^\tc_\g\,,\qquad\qquad
\ki_\tc^\dg{}^\tb{}_\ta\ =\
-\frac{1}{4}\d^\tb_\ta\bla_\tc^\dg\,,\qquad\qquad
\ki_\au{}^\tb{}_\ta\ =\ 0\,. \label{connect} \eea This situation
is analogous to the case of the 16+16 $N=1$ supergravity multiplet
which is obtained from the reducible 20+20 multiplet, described on
superspace with structure group $SL(2,\mathbb{C})\otimes U(4)$, by
``breaking'' the $U(1)$ symmetry \cite{Mul86c}. By eliminating
this $U(1)$ part from the $SL(2,\mathbb{C})\otimes U(4)$
connection and putting the pure gauge part $a$ to zero, one can
define covariant derivatives for $SL(2,\mathbb{C})$ \bea
\hat{D}u^\ca\ =\ Du^\ca-\ki_\cb{}^\ca u^\cb\qquad\quad
\hat{D}u_\ca\ =\ Du^\ca+\ki_\ca{}^\cb u_\cb \eea used in the
articles \cite{GHK01} and \cite{NT81}. Here of course
$\ki_\cb{}^\ca$ is defined in such a way that its only non-zero
components are $\ki^\tb_\b{}^\a_\ta=\d_\b^\a\ki^\tb{}_\ta$ and
$\ki_\tb^\db{}_\da^\ta=-\d^\db_\da\ki^\ta{}_\tb$. Recall that this
redefinition of the connection affects torsion and curvature
components in the following way: \bea \hat{T}_{\cc\cb}{}^\ca & =&
T_{\cc\cb}{}^\ca-\ki_{\cc\cb}{}^\ca+(-)^{cb}\ki_{\cb\cc}{}^\ca,\\
\hat{R}_\cd{}_\cc{}^\tb{}_\ta &=&0\,.
\eea

In the next section we derive the equations of motion for all the
component fields of the N-T multiplet
using its geometrical description presented above.

%%%%%%%%%%%%%%%%%%%%%%%%%%%%%%%%%%%%%%%%%%%%%%%%%%%%%%%%%%%%%%%%%%%%%%
\section{Equations of motion in terms of supercovariant quantities}
%%%%%%%%%%%%%%%%%%%%%%%%%%%%%%%%%%%%%%%%%%%%%%%%%%%%%%%%%%%%%%%%%%%%%%

The problem of the derivation of field equations of motion without
the knowledge of a Lagrangian, using
considerations on representations of the symmetry group, was
considered for a long time \cite{Bel74},
\cite{Wei95}. The question is particularly interesting for
supersymmetric theories and there are various
approaches which have been developed. Let us mention for example the
procedure based on projection
operators selecting irreducible representations out of superfield
with arbitrary external spin
\cite{OS77}. About the same period Wess and Zumino suggested the use
of differential geometry in
superspace to reach better understanding of supersymmetric Yang-Mills
and supergravity theories. The
techniques used in this approach allowed to work out a new method for
deriving equations of motion,
namely looking to consequences of covariant constraints, which
correspond to on-shell field content of a
representation of the supersymmetry algebra.

In order to illustrate the method let us recall as briefly as
possible the simplest example, the $N=1$ Yang-Mills theory
described on superspace considering the geometry of a Lie algebra
valued 1--form $\ca$ \cite{Soh78a}, \cite{BGG01}. Under a gauge
transformation, parameterized by $g$, the gauge potential
transforms as $\ca\mapsto g^{-1}\ca g-g^{-1}\vd g$ and its field
strength $\cf=\vd\ca+\ca\ca$ satisfies the Bianchi identity
$D\cf=0$. In order to describe the off-shell multiplet one
constrains the geometry by putting $\cf_{\a\b} = \cf_\a{}^\db =
\cf^{\da\db} = 0$. Then the Bianchi identities are satisfied if
and only if all the components of the field strength $\cf$ can be
expressed in terms of two spinor superfields $\cw_\a$,
$\bar{\cw}^\da$ and their spinor derivatives: \be \cf_{\b
a}=i(\s_a\bar{\cw})_\b\,,\qquad \cf^\db{}_a=-i(\bs_a\cw)^\db\,,
\ee \be \cf_{(\b\a)}=-\frac{1}{2}\cd_{(\b}\cw_{\a)}\,,\qquad
\cf^{(\db\da)}=\frac{1}{2}\cd^{(\db}\bar{\cw}^{\da)}\,, \ee and
the gaugino superfields $\cw_\a$, $\bar{\cw}^\da$ satisfy \be
\cd_\a\bar{\cw}^\da\ =\ 0\,,\qquad \cd^\da\cw_\a\ =\ 0\,, \eqn{w1}
\be \cd^\a\cw_\a\ =\ \cd_\da\bar{\cw}^\da\,. \ee The components of
the multiplet are thus identified as follows: the vector gauge
field in the super 1--form $\ca\doubar=i\vd x^ma_m$, the gaugino
component field as lowest component of the gaugino superfield
$\cw_\a\loco=-i\la_\a$, $\bar{\cw}^\da\loco=i\la^\da$, and the
auxiliary field in their derivatives
$\cd^\a\cw_\a\loco=\cd_\da\bar{\cw}^\da\loco=-2D$.

Note that the supplementary constraint
\be
\cd^\a\cw_\a\ =\ \cd_\da\bar{\cw}^\da\ =\ 0
\eqn{w2}
  puts this multiplet on-shell. It is a superfield equation and
contains all the component field equations of motion. First of all
it eliminates the auxiliary field $D$ and we can derive the
equations of motion for the remaining fields by successively
differentiating it. We obtain the Dirac equation for the gaugino
\bea
\cd^\da\left(\cd^\a\cw_\a\right)&=& -2i\cd^{\a\da}\cw_\a\ =\ 0\,,\\[2mm]
\cd_\a\left(\cd_\da\bar{\cw}^\da\right)&=& -2i\cd_{\a\da}\bar{\cw}^\da\ =\ 0\,,
\eea
and from this we derive the relations
\bea
\cd_\b\left(\cd^{\a\da}\cw_\a\right)&=&-2\cd^{\a\da}\cf_{(\b\a)}
+2i\{\cw_\b,\bar{\cw}^\da\}\ =\ 0\,,\\[2mm]
\cd^\db\left(\cd_{\a\da}\bar{\cw}^\da\right)&=&2\cd_{\a\da}\cf^{(\db\da)}
-2i\{\bar{\cw}^\db,\cw_\a\}\ =\ 0\,,
\eea
which correspond to the well-known Bianchi identities
$\cd_{\a\db}\cf^{(\db\da)} -
\cd^{\b\da}\cf_{(\b\a)} = 0$ and equations of motion
$\cd_{\a\db}\cf^{(\db\da)} + \cd^{\b\da}\cf_{(\b\a)}
= 2i\{\cw_\a,\bar{\cw}^\db\}$ for the vector gauge field.

The case of supergravity is similar to this, the gravigino
superfields  $T_{[\tc\tb\ta]\a}$,
$T^{[\tc\tb\ta]\da}$ (or $\la_\a^\ta$, $\bla^\da_\ta$ in \equ{Tla})
play an analogous r\^ole to the
gaugino superfields $\cw_\a$, $\bar{\cw}^\da$. In order to derive the
free equations of motion of
component fields in a supergravity theory it is sufficient to
consider only the linearized version
\cite{HL81}, \cite{Sie81b}, \cite{GG83} and the calculations are
simple. Considering the full theory one
obtains all the nonlinear terms which arise in equations of motion
derived from a Lagrangian in component
formalism.

Recall that the dimension 1 Bianchi identities in the supergravity
sector imply the relations
\equ{dim1} for the spinor derivatives of the gravigino superfields.
These properties can be written
equivalently as
\be
\begin{array}{lcl}
\sum_{\td\tc}\cd_\td^\dd T_{[\tc\tb\ta]\a}\ =\ 0\,,&&
\sum^{\td\tc}\cd^\td_\d T^{[\tc\tb\ta]\da}\ =\ 0\,,
\\[2mm]
\cd^{\td\a} T_{[\tc\tb\ta]\a}\ =\ 0\,,&&
\cd_{\td\da} T^{[\tc\tb\ta]\da}\ =\ 0\,,
\\[2mm]
\cd^\td_{(\d}T_{\a)[\tc\tb\ta]}
-\frac{1}{4}\d^{\td\te\tf}_{\tc\tb\ta}\cd^\tg_{(\d}T_{\a)[\tg\te\tf]}\ =\ 0\,,
&&
\cd_\td^{(\dd}T^{\da)[\tc\tb\ta]}
-\frac{1}{4}\d_{\td\te\tf}^{\tc\tb\ta}\cd_\tg^{(\dd}T^{\da)[\tg\te\tf]}\
=\ 0\,,
\end{array}
\ee
and they are the $N=4$ analogues of the relations \equ{w1} and
\equ{w2} satisfied by the gaugino
superfield corresponding to the on-shell Yang-Mills multiplet.

Therefore, by analogy to the Yang--Mills case, the equations of
motion for the gravigini, the
graviphoton, the scalar and the antisymmetric tensor can be deduced
from the superfield relations
\equ{dim1} by taking successive covariant spinorial derivatives.

Consider all possible spinorial derivatives of relations \equ{dim1}.
They are satisfied if and only if in addition to the dimension 1
results the following relations take place:
\bea
\cd_\g^\tc G_{(\b\a)[\tb\ta]}& =& \frac{1}{3}\d^{\tc\tf}_{\tb\ta}\left[
\frac{1}{3}\oint_{\g\b\a}\cd_\g^\te G_{(\b\a)[\te\tf]}
+\frac{i}{2}\sum_{\b\a}\eps_{\g\b}\la_{\da\tf} \bP_\a{}^\da\right]\\
\cd^\dg_\tc G^{(\db\da)[\tb\ta]}& =& \frac{1}{3}\d_{\tc\tf}^{\tb\ta}\left[
\frac{1}{3}\oint^{\dg\db\da}\cd^\dg_\te G^{(\db\da)[\te\tf]}
+\frac{i}{2}\sum^{\db\da}\eps^{\dg\db}\la^{\a\tf} P_\a{}^\da\right]
\eea
\bea
\cd^\tc_\d G^{(\db\da)[\tb\ta]}&=&\cd_\d{}^{(\db}T^{\da)[\tc\tb\ta]}
+U^\tc_\d{}^{(\da}_\tf T^{\db)[\tf\tb\ta]}
-U^\tb_\d{}^{(\da}_\tf T^{\db)[\tf\ta\tc]}
-U^\ta_\d{}^{(\da}_\tf T^{\db)[\tf\tc\tb]}\\
\cd_\tc^\dd G_{(\b\a)[\tb\ta]}&=&\cd_{(\b}{}^{\dd}T_{\a)[\tc\tb\ta]}
-U^\tf_{(\a}{}^{\dd}_\tc T_{\b)[\tf\tb\ta]}
+U^\tf_{(\a}{}^{\dd}_\tb T_{\b)[\tf\ta\tc]}
+U^\tf_{(\a}{}^{\dd}_\ta T_{\b)[\tf\tc\tb]}
\eea
and
\bea
\cd_{\b\da}T^{[\tc\tb\ta]\da}&=&\frac{3}{2}U_\b^\tf{}^\da_\tf
T^{[\tc\tb\ta]}_\da\label{Dirac1}\\
\cd^{\a\db}T_{[\tc\tb\ta]\a}&=&-\frac{3}{2}U_\a^\tf{}^\db_\tf
T_{[\tc\tb\ta]}^\a\label{Dirac2} \eea with $U^\tb_\b{}^\da_\ta =
\frac{i}{4}(\la^\tb_\b\bla^\da_\ta-\frac{1}{2}\d^\tb_\ta\la^\tf_\b\bla^\da_\tf)$.

\vspace{0.5cm}
\noindent{\bf Equations of motion for the helicity 1/2 fields.}

Note first, that all these relations are implied also by Bianchi
identities at dim 3/2. Secondly, note that the last equations,
\equ{Dirac1} and \equ{Dirac2}, are the Dirac equations for the spin
1/2 fields, which may be written in terms of the fields $\la$ in the
following way:
\bea
\cd_{\b\da}\bla^\da_\ta&=&ie^{-2\f}H^*_{\b\da}\bla^\da_\ta+\frac{9i}{8}(\bla_\ta\bla_\tf)\la^\tf_\b\,,\label{Diracc1}\\
\cd^{\a\db}\la_\a^\ta&=&-ie^{-2\f}H^*{}^{\a\db}\la_\a^\ta+\frac{9i}{8}(\la^\ta\la^\tf)\bla_\tf^\db\,.\label{Diracc2}
\eea

\begin{enumerate}[a.]
\item Consider the spinorial derivative $\cd^\td_\d$ of the Dirac
equation \equ{Dirac1}. The derived identity is satisfied if and only
if in addition to the results obtained till dimension 3/2 the
following relations take place:
\be
\cd_{\a\da}P^{\a\da}-i\left(e^{-2\f}H^*_{\a\da}+\frac{1}{2}\la^\tf_\a\bla_{\da\tf}\right)P^{\a\da}
+\frac{iq}{2}\e_{\td\tc\tb\ta}G^{(\db\da)[\td\tc]}G_{(\db\da)}{}^{[\tb\ta]}\
=\ 0
\eqn{dL1}
\be
\sum_{\b\a}\left[\cd_{\b\da}P_\a{}^{\da}-i\left(e^{-2\f}H^*_{\b\da}+\la^\tf_\b\bla_{\da\tf}\right)P_\a{}^{\da}\right]\
=\ 0.
\eqn{dH1}
\item Consider the spinorial derivative $\cd_\td^\dd$ of
\equ{Dirac2}. The identity is satisfied if and only if in addition to
the results obtained till dimension 3/2 the following relations take
place:
\be
\cd_{\a\da}\bP^{\a\da}+i\left(e^{-2\f}H^*_{\a\da}+\frac{1}{2}\la^\tf_\a\bla_{\da\tf}\right)\bP^{\a\da}
+\frac{iq}{2}\e^{\td\tc\tb\ta}G_{(\b\a)[\td\tc]}G^{(\b\a)}{}_{[\tb\ta]}\ =\ 0
\eqn{dL2}
\be
\sum^{\db\da}\left[\cd^{\a\db}\bP_\a{}^{\da}+i\left(e^{-2\f}H^*{}^{\a\db}+\la^{\tf\a}\bla_{\tf}^\db\right)\bP_\a{}^{\da}\right]\
=\ 0.
\eqn{dH2}
\end{enumerate}

\vspace{0.5cm}
\noindent{\bf Equations of motion for the scalar.}

Using properties \equ{PP} the equations of motion for the scalar can be deduced from
the sum of the
relations \equ{dL1} and \equ{dL2}:
\bea
2\cd_a\left(\cd^a
\f\right)&=&e^{-4\f}H^*_aH^{*a}-e^{-2\f}H^*_a(\la^\ta\s^a\bla_\ta)\nn\\
&&-\frac{3}{8}(\la^\tb\la^\ta)(\bla_\tb\bla_\ta)
-\frac{1}{2}e^{-2\f}F_{ba[\tb\ta]}F^{ba[\tb\ta]}\,.\label{scal}
\eea
This equation already shows that in the Lagrangian corresponding to
these equations of motion the kinetic
terms of the antisymmetric tensor and of the graviphotons are
accompanied by exponentials in the scalar
field.

By the way, the difference of relations
\equ{dL1} and
\equ{dL2} looks as
\be
\cd_aH^{*a}\ =\
\frac{1}{2}e^{2\f}(\la^\ta\s_a\bla_\ta)\cd^a\f+\frac{i}{2}F^{*ba[\tb\ta]}F_{ba[\tb\ta]},
\ee
and it corresponds of course to the Bianchi identity satisfied by the
antisymmetric tensor gauge field.
The topological term $F^{*ba[\tb\ta]}F_{ba[\tb\ta]}$ is an indication
of the intrinsic presence of
Chern-Simons forms in the geometry. This feature is analogous to the
case of the off-shell $N=2$ minimal
supergravity multiplet containing an antisymmetric tensor
\cite{AGHH99}. It arises naturally in extended
supergravity using the soldering mechanism with the geometry of a
2--form in central charge superspace.

\vspace{0.5cm}
\noindent{\bf Equations of motion for the antisymmetric tensor.}

Note that relations \equ{dH1} and \equ{dH2} are the selfdual and
respectively the anti-selfdual part of the equation of motion for the
antisymmetric tensor. Putting these relations together, we obtain the
equation of motion for the antisymmetric tensor:
\bea
\e_{dcba}\cd^bH^{*a}&=&\left[T_{dc}{}^\a_\ta\la^\ta_\a+T_{dc}{}_\da^\ta\bla_\ta^\da\right]e^{2\f}
-\frac{1}{2}H^*_{[d}(\la^\tf\s_{c]}\bla_\tf)\nn\\
&&+\e_{dcba}\left[\frac{3}{4}\cd^b(\la^\tf\s^a\bla_\tf)
-(\cd^b\f)(\la^\tf\s^a\bla_\tf)
+4e^{-2\f}(\cd^b \f)H^*{}^a\right]e^{2\f}\label{tens}
\eea

Consider the spinorial derivative $\cd_\td^\dd$ of the Dirac equation
\equ{Dirac1} and the spinorial derivative $\cd^\td_\d$ of
\equ{Dirac2}. The identities obtained this way are satisfied if and
only if in addition to the results obtained till dimension 3/2 the
following relations hold:
\bea
4i\cd_{\b\da}G^{(\dd\da)[\tb\ta]}
&=&q\e^{\tb\ta\td\tc}\left(G_{(\b\a)[\td\tc]}P^{\a\dd}+i\bla^\da_\td\bla^\dd_\tc\bP_{\b\da}\right)\nn\\
&&-G^{(\dd\da)[\tb\tf]}\la^\ta_\b\bla_{\tf\da}
-G^{(\dd\da)[\tf\ta]}\la^\tb_\b\bla_{\tf\da}\label{G1}\\[4mm]
4i\cd^{\a\db}G_{(\d\a)[\tb\ta]}
&=&q\e_{\tb\ta\td\tc}\left(G^{(\db\da)[\td\tc]}\bP_{\d\da}+i\la_\a^\td\la_\d^\tc
P^{\a\db}\right)\nn\\
&&+G_{(\d\a)[\tb\tf]}\la^{\tf\a}\bla_\ta^\db
+G_{(\d\a)[\tf\ta]}\la^{\tf\a}\bla_\tb^\db.\label{G2}
\eea

\vspace{0.5cm}
\noindent{\bf Equations of motion for the graviphotons.}

Recall that the geometric soldering mechanism between supergravity
and the geometry of the 3-form implies
that the fields $G_{(\b\a)[\tb\ta]}$ and $G^{(\db\da)[\tb\ta]}$ are
related to the covariant
field strength of the graviphotons, $F^\au$, by \equ{G-F}. Then the
previous lemma determines both the
equations of motion and the Bianchi identities satisfied by the
vector gauge fields of the multiplet:
\bea
\cd_b
F^{ba\au}&=&-\frac{i}{2}\left[(P_b+\bP_b)F^{ba\au}+(P_b-\bP_b)F^*{}^{ba\au}\right]\nn\\[2mm]
&&
+\frac{i}{4}\left[P_b(\la^\tb\s^{ba}\bla^\ta)\tt_{[\tb\ta]}{}^\au
+\bP_b(\la_\tb\bs^{ba}\bla_\ta)\tt^{[\tb\ta]}{}^\au\right]e^\f\nn\\[2mm]
&&-\frac{i}{4}\left[(\la^\ta\s^{dc}\s^a\bla_\tb)F_{dc}{}^\av\hh_\av{}^{[\tb\tf]}\tt_{[\tf\ta]}{}^\au
+(\bla_\ta\bs^{dc}\bs^a\la^\tb)F_{dc}{}^\av\hh_\av{}_{[\tb\tf]}\tt^{[\tf\ta]}{}^\au\right]\,,\label{F}\\[2mm]
\cd_b F^*{}^{ba\au}&=&
\frac{i}{4}\left[P_b(\la^\tb\s^{ba}\bla^\ta)\tt_{[\tb\ta]}{}^\au
-\bP_b(\la_\tb\bs^{ba}\bla_\ta)\tt^{[\tb\ta]}{}^\au\right]e^\f\nn\\[2mm]
&&-\frac{i}{4}\left[(\la^\ta\s^{dc}\s^a\bla_\tb)F_{dc}{}^\av\hh_\av{}^{[\tb\tf]}\tt_{[\tf\ta]}{}^\au
-(\bla_\ta\bs^{dc}\bs^a\la^\tb)F_{dc}{}^\av\hh_\av{}_{[\tb\tf]}\tt^{[\tf\ta]}{}^\au\right]\,.\label{F*}
\eea

Further differentiating \equ{G1} and \equ{G2} one can obtain Bianchi
identities for the gravitini and graviton, but here we would like to
derive their equations of motion instead.

\vspace{0.5cm}
\noindent{\bf Equations of motion for the gravitini.}

Unlike the equations of motion presented above, the equations of
motion for the gravitini and the graviton are directly given by the
superspace Bianchi identities, once the component fields are
identified. For example, the Bianchi identities at dim 3/2
determinate the torsion components
\be
T_{(\b\a)}{}^\a_\ta\ =\ \frac{1}{16}\bP_{\b\da}\bla^\da_\ta
\qquad T^{(\dd\dg)}{}_{\b\td}\ =\ \frac{1}{8}\bP_{\b}{}^{(\dd}\bla^{\dg)}_\td
+\frac{iq}{8}\e_{\td\tc\tb\ta}G^{(\dd\dg)[\tc\tb]}\la^\ta_\b\,,
\ee
\be
T^{(\db\da)}{}_\da^\ta\ =\ \frac{1}{16}P^{\a\db}\la_\a^\ta
\qquad T_{(\d\g)}{}^{\db\td}\ =\ \frac{1}{8}P_{(\d}{}^{\db}\la_{\g)}^\td
+\frac{iq}{8}\e^{\td\tc\tb\ta}G_{(\d\g)[\tc\tb]}\bla_\ta^\db\,,
\ee
and these components are sufficient to give the equations of motion
for the gravitini:
\be
\e^{dcba}(\bs_cT_{ba\ta})^\da\ =\ \frac{i}{4}(\bla_\ta\bs^d\s^c\eps)^\da\bP_c
+\frac{i}{2}(\bs^{ba}\bs^d\la^\tf)^\da F_{ba}{}_{[\ta\tf]}e^{-\f}\,,
\eqn{gravitini1}
\be
\e^{dcba}(\s_cT_{ba}{}^\ta)_\a\ =\ -\frac{i}{4}(\la^\ta\s^d\bs^c\eps)_\a P_c
-\frac{i}{2}(\s^{ba}\s^d\bla_\tf)_\a F_{ba}{}^{[\ta\tf]} e^{-\f}\,.
\eqn{gravitini2}

\vspace{0.5cm}
\noindent{\bf Equations of motion for the graviton.}

In order to give the equations of motion for the graviton we need the
expression of the supercovariant
Ricci tensor, $R_{db}=R_{dcba}\eta^{ca}$, which is given by the
superspace Bianchi identities at
canonical dimension 2 \equ{Riccitensor}. The corresponding Ricci
scalar, $R=R_{db}\eta^{db}$, is then
\be
R=-2\cd^a\f\cd_a\f-\frac{1}{2}H^*{}^aH^*{}_ae^{-4\f}
+\frac{3}{4}e^{-2\f}H^*{}^a(\la^\ta\s_a\bla_\ta)
+\frac{3}{8}(\la^\tb\la^\ta)(\bla_\tb\bla_\ta).
\ee
The knowledge of these ingredients allows us to write down the
Einstein equation
\bea
R_{db}-\frac{1}{2}\eta_{db}R&=&-2\left[\cd_d\f\,\cd_b\f-\frac{1}{2}\eta_{db}\cd^a\f\cd_a\f\right]\nn\\[2mm]
&&-\frac{1}{2}e^{-4\f}\left[H^*_d\,H^*_b-\frac{1}{2}\eta_{db}H^*{}^aH^*{}_a\right]\nn\\[2mm]
&&-e^{-2\f}\left[F_{df[\tb\ta]}F_b{}^f{}^{[\tb\ta]}-\frac{1}{4}\eta_{db}F_{ef[\tb\ta]}F^{ef[\tb\ta]}\right]\nn\\[2mm]
&&-\frac{i}{8}\sum_{db}\left[\la^\tf\s_d\cd_b\bla_\tf-(\cd_b\la^\tf)\s_d\bla_\tf\right]\nn\\[2mm]
&&-\frac{1}{8}\left[\frac{1}{4}(\la^\tf\s_d\bla_\tf)\,(\la^\ta\s_b\bla_\ta)+\eta_{db}(\la^\tb\la^\ta)(\bla_\tb\bla_\ta)\right]\nn\\[2mm]
&&-\frac{1}{8}e^{-2\f}\left[H_d^*(\la^\tf\s_b\bla_\tf)-3\eta_{db}H^*{}^a(\la^\ta\s_a\bla_\ta)\right]\,,
\label{gr}
\eea
where one may recognize on the right-hand-side the usual terms of the
energy-momentum tensor
corresponding to matter fields: scalar fields, antisymmetric tensor,
photon fields and spinor fields
respectively. As it will be shown by \equ{Rdbbar}, the contribution
of the gravitini is hidden in
$R_{db}$.

%%%%%%%%%%%%%%%%%%%%%%%%%%%%%%%%%%%%%%%%%%%%%%%%%%%%%%%%%%%%%
\section{Equations of motion in terms of component fields}
%%%%%%%%%%%%%%%%%%%%%%%%%%%%%%%%%%%%%%%%%%%%%%%%%%%%%%%%%%%%%

In the previous section we calculated the equations of motion for all
component fields of the N-T
multiplet (graviton \equ{gr}, gravitini \equ{gravitini1},
\equ{gravitini2},  graviphotons \equ{F},
1/2-spin fields \equ{Diracc1}, \equ{Diracc2}, scalar \equ{scal} and
the antisymmetric tensor \equ{tens})
in terms of supercovariant objects, which have only flat (Lorentz)
indices. In order to write these
equations of motion in terms of component fields, one passes to
curved (Einstein) indices by the standard
way \cite{WB83}. General formulas are easily written using the
notation $E^\ca\doubar=e^\ca=\vd x^m
e_m{}^\ca$ \cite{BGG01}.

%=======================================================
\subsection{Supercovariant$\rightarrow$component toolkit}
%=======================================================

Recall that the graviton, gravitini and graviphotons are identified
in the super-vielbein. Thus, their
field strength can be found in their covariant counterparts using
\be
T^\ca\doubar\  =\  \frac{1}{2}\vd x^m \vd x^n
\left(\cd_ne_m{}^\ca-\cd_me_n{}^\ca\right)
\  =\  \frac{1}{2}e^\cb e^\cc T_{\cc\cb}{}^\ca\loco\,.
\ee
For $\ca=a$ one finds the relation
\be
\cd_ne_m{}^a-\cd_m e_n{}^a\ =\ i\p_{[n\ta}\s^a\bar{\p}_{m]}{}^\ta\,,
\eqn{Tcba}
which determinates the Lorentz connection in terms of the vierbein,
its derivatives and gravitini fields.
For $\ca=_\a^\ta$ and $\ca=^\da_\ta$ we have the expression of the
covariant field strength of the
gravitini
\bea
T_{cb}{}_\ta^\a\loco &=& e_b{}^me_c{}^n\cd_{[n}\p_{m]}{}^\a_\ta%T_{nm}{}_\ta^\a
-e_b{}^me_c{}^n\frac{q}{4}\e_{\td\tc\tb\ta}
\bar{\p}_n{}^{\td}\bar{\p}_m{}^{\tc}\la^{\a}{}^\tb
-\frac{i}{2}e_{[c}{}^n(\bar{\p}_n{}^{\tb}\bar{\s}_{b]}\s^{da})^{\a}F_{da}{}_{[\tb\ta]}\loco
e^{-\f}\nn\\[2mm]
&&
+\frac{i}{4}(\p_{n\tf}\s_{f[b})^{\a}e_{c]}{}^n
\left[\la^{\tf}\s^f\bar{\la}_{\ta}
-\frac{1}{2}\d_{\ta}^{\tf}\la^{\tb}\s^f\bar{\la}_{\tb}\right]\,,
\eea
\bea
T_{cb}{}^\ta_\da\loco &=&
e_b{}^me_c{}^n\cd_{[n}\bp_{m]}{}_\da^\ta%T_{nm}{}^\ta_\da
-e_b{}^me_c{}^n\frac{q}{4}\e^{\td\tc\tb\ta}
{\p}_n{}_{\td}{\p}_m{}_{\tc}\bar{\la}_{\da}{}_\tb
-\frac{i}{2}e_{[c}{}^n({\p}_n{}_{\tb}{\s}_{b]}\bs^{da})_{\da}F_{da}{}^{[\tb\ta]}\loco
e^{-\f}\nn\\[2mm]
&&
-\frac{i}{4}(\bar{\p}_n{}^\tf\bar{\s}_{f[b})_{\da}e_{c]}{}^n
\left[{\la}^{\ta}\s^f\bar{\la}_{\tf}
-\frac{1}{2}\d^{\ta}_{\tf}\la^{\tb}\s^f\bar{\la}_{\tb}\right]\,.
\eea
As for $\ca=\au$, the central charge indices, we obtain the covariant
field strength of the graviphotons
\bea
F_{ba}{}^\au\loco &=& e_b{}^n e_a{}^m\cf_{nm}{}^\au%2\partial_{[n}v_{m]}{}^\au
+e_b{}^n e_a{}^m\left[\bar{\p}_{n}{}^{\tc}\bar{\p}_{m}{}^{\tb}
+i\bar{\p}_{[n}{}^{\tc}\bar{\s}_{m]}\la^{\tb}\right]e^\f\tt_{[\tc\tb]}{}^{\au}\nn\\[2mm]
&&
+e_b{}^n
e_a{}^m\left[\p_{n\tc}\p_{m\tb}+i\p_{[n\tc}\s_{m]}\bar{\la}_{\tb}\right]e^\f\tt^{[\tc\tb]\au}\,,
\eea
with $\cf_{nm}{}^\au$ the field strength of the graviphotons
$\cf_{nm}{}^\au=\partial_{n}v_{m}{}^\au
-\partial_{m}v_{n}{}^\au$. In the $SU(4)$ basis this becomes
\bea
F_{ba}{}^{[\tb\ta]}\loco &=& e_b{}^n e_a{}^m\cf_{nm}{}^{[\tb\ta]}
%F_{nm}{}^{[\tb\ta]}
+e_b{}^n e_a{}^m\left[\bar{\p}_{n}{}^{[\tb}\bar{\p}_{m}{}^{\ta]}
+i\bar{\p}_{[n}{}^{[\tb}\bar{\s}_{m]}\la^{\ta]}\right]e^\f\nn\\[2mm]
&&
+e_b{}^n e_a{}^m\frac{q}{2}\e^{\td\tc\tb\ta}\left[\p_{n\td}\p_{m\tc}
+i\p_{[n\td}\s_{m]}\bar{\la}_{\tc}\right]e^\f\,,
\eea
with the field
strength
$\cf_{nm}{}^{[\tb\ta]}=\cf_{nm}{}^\au\hh_\au{}^{[\tb\ta]}=\partial_{n}V_{m}{}^{[\tb\ta]}
-\partial_{m}V_{n}{}^{[\tb\ta]}$.

Since the antisymmetric tensor is identified in the 2--form, the
development of its covariant field
strength on component fields is deduced using
\be
H\doubar\ =\ \frac{1}{2}\vd x^m\vd x^n\vd x^k \partial_kb_{nm}\ =\
\frac{1}{3!}e^\ca e^\cb e^\cc H_{\cc\cb\ca}\loco
\ee
and one finds
\bea
H^*{}^a\loco &=& e_l{}^a\cg^l
+ie_l{}^a\left[\p_{k\tf}\s^{lk}\la^{\tf}
-\bar{\p}_k{}^{\tf}\bar{\s}^{lk}\bar{\la}_{\tf}
+\frac{1}{2}\e^{lknm}\p_{k\tf}\s_n\bar{\p}_{m}{}^{\tf}\right]e^{2\f}\,,
\label{compH}
\eea
with
\be
\cg^l\ =\
\frac{1}{2}\e^{lknm}\left[\partial_kb_{nm}-v_k{}^\au\gg_{\au\av}\cf_{nm}{}^\av\right]
\ =\
\frac{1}{2}\e^{lknm}\left[\partial_kb_{nm}-V_{k[\tb\ta]}\cf_{nm}{}^{[\tb\ta]}\right]\,.
\ee
Note, that the dual field strength,
$\frac{1}{2}\e^{lknm}\partial_kb_{nm}$, of the antisymmetric tensor
appears in company with the Chern-Simons term
$\frac{1}{2}\e^{lknm}v_k{}^\au\gg_{\au\av}\cf_{nm}{}^\av%=\frac{1}{2}\e^{lknm}V_{k[\tb\ta]}\cf_{nm}{}^{[\tb\ta]}
$. We use the notation $\cg^l$ in order to accentuate this feature.
Recall also, that one of the
fundamental aims of the article \cite{GHK01} was to explain in detail
that this phenomenon is quite
general and arises as an intrinsic property of soldering in
superspace with central charge coordinates.

The lowest component of the derivative of the scalar can be
calculated using $D\f\doubar = \vd x^m\cd_m\f
= e^\ca\cd_\ca \f\loco$, and it is
\be
\cd_a\f\loco = e_a{}^m\left(\cd_m\f
-\frac{1}{4}\p_{m\tf}\la^{\tf}-\frac{1}{4}\bar{\p}_m{}^{\tf}\bar{\la}_{\tf}\right)\,,
\ee
while the lowest component of the double derivative
$\cd_a\cd^a\f\loco$, needed for the expansion of the
equation of motion for the scalar \equ{scal}, becomes
\bea
2\cd_a\cd^a\f\loco &=& 2\Box \f
+e_a{}^m\cd_me^{an}\left[2\cd_n\f
-\frac{1}{2}\p_{n\tf}\la^{\tf}-\frac{1}{2}\bar{\p}_n{}^{\tf}\bar{\la}_{\tf}\right]
-\frac{1}{2}H^{*a}\loco\p_{m\tf}\s_a\bar{\p}^{m\tf}\, e^{-2\f}\nn\\[2mm]
&&
-\frac{1}{2}\cd^m(\p_{m\tf}\la^{\tf}+\bar{\p}_m{}^{\tf}\bar{\la}_{\tf})
-\frac{1}{2}(\p_{m\tf}\cd^m\la^{\tf}+\bar{\p}_m{}^{\tf}\cd^m\bar{\la}_{\tf})\nn\\[2mm]
&&
-\frac{3i}{32}(\la^{\tc}\s^n\bar{\la}_{\tc})\left(\p_{n\tf}\la^{\tf}-\bar{\p}_n{}^{\tf}\bar{\la}_{\tf}\right)
-\frac{3}{4}(\p_{m\tf}\la^{\ta})(\bar{\p}_m{}^{\tf}\bar{\la}_{\ta})\nn\\[2mm]
&&
-\frac{1}{4}(\p_{m\tf}\la^{\tf})(\p^m_{\tc}\la^{\tc})
+\frac{1}{2}(\p^m_{\tf}\la^{\tf})(\bar{\p}_m{}^{\tc}\bar{\la}_{\tc})
-\frac{1}{4}(\bar{\p}_m{}^{\tf}\bar{\la}_{\tf})(\bar{\p}^{m\tc}\bar{\la}_{\tc})\nn\\[2mm]
&&
+F_{ba[\tf\tc]}\loco\left[\frac{1}{2}\bar{\p}_m{}^{\tf}\bar{\s}^{ba}\bar{\p}^{m\tc}
+\frac{i}{4}\bar{\p}_n{}^{\tf}\bar{\s}^n\s^{ba}\la^{\tc}\right]e^{-\f}\nn\\[2mm]
&&
+F_{ba}{}^{[\tf\tc]}\loco\left[\frac{1}{2}\p_{m\tf}\s^{ba}\p^m_{\tc}
+\frac{i}{4}\p_n{}_{\tf}\s^n\bar{\s}^{ba}\bar{\la}_{\tc}\right]e^{-\f}\,.
\eea
In order to compare our results with the component expression of the
scalar's equation of motion derived
from \cite{NT81}, we have to replace in this expression
$e_a{}^m\cd_me^{an}$ with
\[
e_a{}^m\cd_me^{an}\ =\
V^{-1}\partial_m(Vg^{mn})-ig^{mn}\p_{[m\ta}\s^k\bp_{k]}{}^\ta\,,
\]
a consequence of \equ{Tcba}.

Finally, using $R_b{}^a\doubar = \frac{1}{2}\vd x^m\vd x^n \cR_{nmb}{}^a =
\frac{1}{2}e^\cb e^\cc R_{\cc\cb}{}_b{}^a\loco $, one obtains for the lowest
component of the covariant Ricci tensor $ R_{db}$ the expression

\bea
R_{db}\loco &=& \frac{1}{2}\sum_{db}\left\{e_d{}^ne^{ma}{\cal R}_{nmba}\loco
+\frac{1}{2}\e_b{}^{mef}\p_{m\td}\s_dT_{ef}{}^\td\loco
-\frac{1}{2}\e_b{}^{mef}\bp_m{}^{\td}\bs_d T_{ef\,\td}\loco\right.\nn\\[2mm]
&&+\frac{1}{4}\left(i\p_\td^m\s_d\bs^f\s_{bm}
\la^\td-ie_d{}^n\d_b^f\p_{n\td}\la^\td\right)P_f\loco\nn\\[2mm]
&&+\frac{1}{4}\left(i\bp^{m\td}\bs_d\s^f\bs_{bm}
\bla_\td-ie_d{}^n\d_b^f\bp_n^\td\bla_\td\right)\bar{P}_f\loco\nn\\[2mm]
&&-\frac{1}{2}e^{-\f}F^{ef[\td\tf]}\loco\left(i\textrm{tr}(\s_{bm}\s_{ef})\p_\td^m\s_d\bla_\tf+\frac{i}{2}
e_d{}^n\p_{n\td}\s_{ef}\s_b\bla_\tf\right)
\nn\\[2mm]
&&-\frac{1}{2}e^{-\f}F^{ef}{}_{[\td\tf]}\loco\left(i\textrm{tr}(\bs_{bm}\bs_{ef})\bp^{m\td}\bs_d\la^\tf+\frac{i}{2}
e_d{}^n\bp_n^\td\bs_{ef}\bs_b\la^\tf\right)
\nn\\[2mm]
&&+\frac{1}{2}e_d{}^ne^{-\f}\left(\textrm{tr}(\bs_b{}^m\bs_{ef})\p_{n\td}\p_{m\tc}F^{ef[\td\tc]}\loco
+\textrm{tr}(\s_b{}^m\s_{ef})\bp_n^\td\bp_m^\tc
F^{ef}{}_{[\td\tc]}\loco\right)\nn\\[2mm]
&&-\left.\frac{1}{2}e_d{}^m(\d^\td_\tb\d^\tc_\ta-\frac{1}{2}\d^\td_\ta\d^\tc_\tb)
\left[(\p_{[m\td}\s_b{}^n\la^\tb)(\bp_{n]}{}^\ta\bla_\tb)-(\p_{[m\td}\la^\tb)(\bp_{n]}{}^\ta\bs_b{}^n\bla_\tb)
\right]\right\}.
\label{Rdbbar}
\eea

%==========================================
\subsection{The equations of motion}
%==========================================

In the last subsection we deduced the expression of all quantities
appearing in the
supercovariant equations of motion in terms of component fields. We
are therefore
ready now to replace these expressions in \equ{gr}, \equ{gravitini1},
\equ{gravitini2}, \equ{F}, \equ{Diracc1}, \equ{Diracc2}, \equ{scal}, \equ{tens}
  and give the equations of motion in terms
of component fields.

It turns out that the expressions
\bea \tilde{H}_{l} &=& e_l{}^a H^*_a\loco
-\frac{i}{2}e^{2\f}\p_{l\ta}\la^{\ta}
+\frac{i}{2}e^{2\f}\bp_l{}^{\ta} \bla_{\ta}
-\frac{3}{4}e^{2\f}\la^{\ta}\s_l\bla_{\ta}\nn\\[2mm]
&=&\cg_l +\frac{i}{2}e^{2\f}\left[\p_{k\tf}\s_l\bs^{k}\la^{\tf}
-\bar{\p}_k{}^{\tf}\bs_l\s^{k}\bar{\la}_{\tf}
+\e_l{}^{knm}\p_{k\tf}\s_n\bar{\p}_{m}{}^{\tf}\right]
-\frac{3}{4}e^{2\f}\la^{\ta}\s_l\bla_{\ta}
\eea
and
\bea
\tilde{F}^{nk\az} &=&e^n{}^be^k{}^a F_{ba}{}^\az\loco
+\frac{i}{2}\e^{nkml}
\left[\bp_m{}^{\td}\bp_l{}^{\tc}
-i\frac{q}{2}\e^{\td\tc\tb\ta}\bla_{\tb}\bs_m\p_{l\ta}\right]e^{\f}t_{[\td\tc]}{}^{\az}\nn\\[2mm]
&&-\frac{i}{2}\e^{nkml}
\left[\p_m{}_{\td}\p_l{}_{\tc}
-i\frac{q}{2}\e_{\td\tc\tb\ta}\la^{\tb}\s_m\bp_l{}^\ta\right]e^{\f}t^{[\td\tc]}{}^{\az}\nn\\[2mm]
&=&
\cf^{nk\az}-\textrm{tr}(\s^{nk}\s^{ml})
\left[\bp_m{}^{\td}\bp_l{}^{\tc}
-i\frac{q}{2}\e^{\td\tc\tb\ta}\bla_{\tb}\bs_m\p_{l\ta}\right]e^{\f}t_{[\td\tc]}{}^{\az}\nn\\[2mm]
&&-\textrm{tr}(\bs^{nk}\bs^{ml})
\left[\p_m{}_{\td}\p_l{}_{\tc}
-i\frac{q}{2}\e_{\td\tc\tb\ta}\la^{\tb}\s_m\bp_l{}^\ta\right]e^{\f}t^{[\td\tc]}{}^{\az}
\eea
appear systematically, and using them, the equations take a quite
simple form. Let us also denote the
quantity $\hat{F}_{nk}{}^\az=e_n{}^be_k{}^a F_{ba}{}^\az\loco$, which
is called the supercovariant field
strength of the graviphotons in the component approach  \cite{CS77b},
\cite{NT81}.

\vspace{0.5cm}
\noindent{\bf Equations of motion for the helicity 1/2 fields.}
\bea
\left(\s^m\hat{\cd}_m\bla_{\ta}\right)_{\b} &=&
-ie^{-2\f}\tilde{H}^m
\left[\frac{i}{2}(\s^n\bar{\s}_m\psi_{n\ta})_{\b}-\frac{3}{4}(\s_m\bla_{\ta})_{\b}\right]
-\frac{i}{2}(\bar{\psi}_n{}^{\tf}\bla_\tf)(\s^m\bs^n\p_{m\ta})_\b\nn\\[2mm]
&&+i\partial_n\f(\s^m\bar{\s}^n\psi_{m\ta})_{\b}-e^{-\f}\hat{F}_{kl[\tf\ta]}(\s^m\bar{\s}^{kl}\bar{\psi}_m{}^{\tf})_{\b}
-\frac{3i}{8}(\bar{\la}_{\ta}\bar{\la}_{\tf}) \la_{\b}^{\tf}
\eea

\vspace{0.5cm}
\noindent{\bf Equations of motion for the gravitini.}
\bea
\e^{lknm}(\bs_k{}\hat{\cd}_n\p_{m\ta})^\da &=&
-\frac{i}{4}e^{-2\f}\tilde{H}_n\left[\e^{lknm}(\bs_k\p_{m\ta})^{\da}
+(\bla_{\ta}\bs^l\s^n\e)^{\da}\right]
-\frac{1}{2}\partial_n\f(\bla_{\ta}\bs^l\s^n\e)^{\da}\nn\\[2mm]
&&-e^{-\f}\hat{F}_{mn[\ta\tf]}\left[\textrm{tr}(\s^{lk}\s^{mn})\bp_k{}^{\tf\da}
+\frac{i}{2}\textrm{tr}(\bs^{lk}\bs^{mn})(\s_k\la^\tf)^\da\right]\nn\\[2mm]
&&+\frac{1}{8}\p_{n\ta}\la^\tf(\bs^{ln}\bla_\tf)^\da+\frac{3}{8}(\p_{n\ta}\s^{ln}\la^\tf)
\bla_\tf^\da-\frac{1}{4}(\p_{n\tf}\s^l\bs^n\la^\tf)\bla_\ta^\da
\nn\\[2mm]
&&+\frac{q}{4}\e^{lknm}\e_{\tc\tb\tf\ta}
\bp_n{}^{\tc}\bp_m{}^{\tb}(\bs_k\la^{\tf})^{\da}
\eea

\vspace{0.5cm}
\noindent{\bf Equations of motion for the scalar.}
\bea
0 &=& 2V^{-1}\partial_m(Vg^{mn}\partial_n\f)
+\frac{1}{2}V^{-1}\partial_m\left(V\la^{\ta}\s^n\bs^m\p_{n\ta}+V\bla_{\ta}\bs^n\s^m\bp_n^{\ta}
\right)\nn\\[2mm]
&&-e^{-4\f}\cg^m\tilde{H}_m
+\frac{1}{2}e^{-2\f}\cf_{nm[\tb\ta]}\tilde{F}^{nm[\tb\ta]}
\eea

\vspace{0.5cm}
\noindent{\bf Equations of motion for the antisymmetric tensor.}
\bea
\partial_{k}\left(e^{-4\f}V\e^{mnkl}\tilde{H}_{l}\right) &=& 0
\eea

\vspace{0.5cm}
\noindent{\bf Equations of motion for the graviphotons.}
\bea
\partial_n\left(Ve^{-2\f}\tilde{F}^{nk\au}\right) &=&
\frac{1}{2}Ve^{-4\f}\e^{lnmk}\tilde{H}_l\cf_{nm}{}^{\au}
\eea

\vspace{0.5cm}
\noindent{\bf Equations of motion for the graviton.}

The Einstein equation in terms of component fields is also deduced in
a straightforward manner from
\equ{gr} and \equ{Rdbbar} with the usual Ricci tensor
$\cR_{mn}=\frac{1}{2}\sum_{mn}e_n{}^be^{ka}\cR_{mkba}$. Here we give
the expression of the Ricci scalar:
\bea \cR &=&
\frac{1}{2}\e^{lknm}\p_{l\ta}\s_k\hat{\cd}_n\bp_m{}^\ta-\frac{1}{2}\e^{lknm}
\bp_l{}^{\ta}\bs_k\hat{\cd}_n\p_{m\ta}\nn\\[2mm]
&&-\frac{i}{4}\la^\ta\s^m\hat{\cd}_m\bla_\ta-\frac{i}{4}\bla_\ta\bs^m\hat{\cd}_m\la^\ta
-2\partial^m\f\partial_m\f\nn\\[2mm]
&&-e^{-\f}\hat{F}_{kl}{}^{[\tb\ta]}\left(\textrm{tr}(\bs^{kl}\bs^{mn})\p_{m\tb}\p_{n\ta}
+\frac{i}{2}\textrm{tr}(\s^{kl}\s^{mn})\p_{m\tb}\s_n\bla_\ta\right)\nn\\[2mm]
&&-e^{-\f}\hat{F}_{kl[\tb\ta]}\left(\textrm{tr}(\s^{kl}\s^{mn})\bp_m{}^\tb\bp_n{}^\ta
+\frac{i}{2}\textrm{tr}(\bs^{kl}\bs^{mn})\bp_m{}^{\tb}\bs_n\la^\ta\right)\nn\\[2mm]
&&-\frac{1}{2}e^{-4\f}\tilde{H}^{l}\left(\cg_l+\frac{i}{2}e^{2\f}(\p_{k\tf}\s_l\bs^{k}\la^{\tf}
-\bar{\p}_k{}^{\tf}\bs_l\s^{k}\bar{\la}_{\tf}
+2\e_l{}^{knm}\p_{k\tf}\s_n\bar{\p}_{m}{}^{\tf}\right)\nn\\[2mm]
&&+\frac{1}{2}(\p_{l\ta}\la^\ta)(\bp^{l\ta}\bla_\ta)+\frac{i}{2}(\la^\tf\s^l\bla_\tf)
(\p_{l\ta}\la^\ta-\bp_l{}^{\ta}\bla_\ta)\nn\\[2mm]
&&-\frac{3i}{16}\e^{lmnk}(\p_{l\tf}\s_m\bp_n{}^\tf)(\la^\ta\s_k\bla_\ta)
-\frac{i}{2}\e^{lmnk}(\p_{l\tf}\s_m\bp_n{}^\ta)(\la^\tf\s_k\bla_\ta) \eea

\section{Conclusion}

The aim of this article was to deduce the equations of motion for the
components of the N-T multiplet
from its geometrical description in central charge superspace, and
compare these equations with those,
deduced from the Lagrangian of the component formulation of the
theory with the same field content
\cite{NT81}.

We showed that the constraints on the superspace which allow to
identify the components in the geometry
imply equations of motion in terms of supercovariant quantities.
Moreover, we succeeded in writing these
equations of motion in terms of component fields in an elegant way,
using the objects $\tilde{H}_m$ and
$\tilde{F}_{mn}{}^\au$. The equations found this way are in perfect
concordance with the ones deduced
from the Lagrangian of Nicolai and Townsend \cite{NT81}. This result
resolves all remaining doubt about
the equivalence of the geometric description on central charge
superspace of the N-T multiplet and the
Lagrangian formulation of the theory with the same field content.

As a completion of this work one may ask oneself about an
interpretation of the objects $\tilde{H}_m$ and
$\tilde{F}_{mn}{}^\au$, which seem to be some natural building blocks
of the Lagrangian. Concerning this
question let us just remark the simplicity of the relation
\be
-i\ki_m{}^\ta{}_\ta\ =\ e^{-2\f}\tilde{H}_m+\frac{3}{8}\la^\ta\s_m\bla_\ta
\ee
between $\tilde{H}_m$ and the $U(1)$ part of the initial connection
\equ{connect} of the central charge
superspace with structure group $SL(2,\mathbb{C})\otimes U(4)$.

\acknowledgments{We thank Richard Grimm for the numerous discussions
on the subject and for his constant support.}

\begin{appendix}
%%%%%%%%%%%%%%%%%%%%%%%%%%%%%%%%%%%%%%%%%%%%%%%%%%%%%%%%%%%%%%%%%%%%%%%%%%%%%%%%%%%%%%%
\section{Solution of the Bianchi Identities}
%%%%%%%%%%%%%%%%%%%%%%%%%%%%%%%%%%%%%%%%%%%%%%%%%%%%%%%%%%%%%%%%%%%%%%%%%%%%%%%%%%%%%%%

The conventional constraints compatible with the assumptions
(\ref{T01} -- \ref{T02}) and \equ{tz} are
the following: \be \begin{array}{lcl} T^\tc_\g{}_b{}^a\ =\
0&\qquad\quad&T_\tc^\dg{}_b{}^a\ =\ 0 \\ [2mm]
T^\tc_\g{}^\tb_\b{}_{\ta}^\a\ =\ 0&&T_\tc^\dg{}_\tb^\db{}^{\ta}_\da\
=\ 0 \\ [2mm]
T^\tc_\g{}^\db_\tb{}_\da^\ta\ =\ 0&&T_\tc^\g{}_\db^\tb{}^\da_\ta\ =\
0 \\ [2mm] T_c{}^\tb_\a{}^\a_\ta\ =\
0&&T_c{}_\tb^\da{}_\da^\ta\ =\ 0 \\ [2mm] &T_{cb}{}^a\ =\ 0\,.&
\end{array} \eqn{conv}

There is a particular solution of the Bianchi identities for the
torsion and 3--form subject to the
constraints (\ref{T01} -- \ref{H02}), (\ref{dt} -- \ref{tz}) and
\equ{conv}, which describes the N-T
supergravity multiplet. Besides the constant $T^\dg_\tc{}^\tb_\b{}^a$
and the supercovariant field
strength of the graviphotons, $T_{cb}{}^\au\doteq F_{cb}{}^\au$, the
non-zero torsion components
corresponding to this solution are then the following:
\be
\begin{array}{lcl}
T_\g^\tc{}_\b^\tb{}^\au\ =\ 4\eps_{\g\b}\tt^{[\tc\tb]}{}^\au e^\f&&
T^\dg_\tc{}^\db_\tb{}^\au\ =\ 4\eps^{\dg\db}\tt_{[\tc\tb]}{}^\au e^\f
\\[2mm]
T^\tc_\g{}^\tb_\b{}^\ta_\da\ =\ q\eps_{\g\b}\e^{\tc\tb\ta\tf}\bla_\tf{}_\da
&\qquad& T_\tc^\dg{}_\tb^\db{}_\ta^\a\ =\
q\eps^{\dg\db}\e_{\tc\tb\ta\tf}\la^\tf{}^\a
\\ [2mm]
T^\tc_\g{}_b{}^\au\ =\ ie^\f(\s_b\bla_\ta)_\g\tt^{[\ta\tc]\au}&&
T_\tc^\dg{}_b{}^\au
\ =\ ie^\f(\bs_b\la^\ta)^\dg\tt_{[\ta\tc]}{}^\au
\\ [2mm]
T^\tc_\g{}_b{}^\a_\ta\ =\ -2(\s_{ba})_\g{}^\a
U^a{}^\tc{}_\ta&&T_\tc^\dg{}_b{}_\da^\ta
\ =\ 2(\bs_{ba})^\dg{}_\da U^a{}^\ta{}_\tc
\\ [2mm]
T^\tc_\g{}_b{}_\da^\ta\ =\
\frac{i}{2}(\s_b\bs^{dc})_{\g\da}F_{dc}{}^{[\tc\ta]}e^{-\f}
&&T_\tc^\dg{}_b{}^\a_\ta\ =\
\frac{i}{2}(\bs_b\s^{dc})^{\dg\a}F_{dc}{}_{[\tc\ta]}e^{-\f}
\\ [2mm]
T_{cb}{}_{\ta\a}\ =\ -(\eps\s_{cb})^{\g\b}\S_{(\g\b\a)\ta}
&& T_{cb}{}^{\ta\da}\ =\ -(\eps\bs_{cb})_{\dg\db}\S^{(\dg\db\da)\ta}
\\ [2mm]
\quad-\frac{1}{4}\textrm{tr}(\bs_{cb}\bs_{af})F^{af}{}_{[\ta\tf]}\la^\tf_\a &&
\quad-\frac{1}{4}\textrm{tr}(\s_{cb}\s_{af})F^{af}{}^{[\ta\tf]}\bla_\tf^\da
\\ [2mm]
\quad-\frac{1}{12}\left(\d^{af}_{cb}-\frac{i}{2}\e_{cb}{}^{af}\right)\bP_a(\s_f\bla_\ta)_\a&&
\quad-\frac{1}{12}\left(\d^{af}_{cb}+\frac{i}{2}\e_{cb}{}^{af}\right)P_a(\bs_f\la^\ta)^\a
\end{array}
\ee
with
$U_a{}^\tb{}_\ta=-\frac{i}{8}(\la^\tb\s_a\bla_\ta-\frac{1}{2}\d^\tb_\ta\la^\tf\s_a\bla_\tf)$,
$P$
and $\bP$ are given in equations \equ{P} and \equ{barP}, while
$\S_{(\g\b\a)\ta}$ and
$\S^{(\dg\db\da)\ta}$ are the gravitino "Weyl" tensors.

Furthermore, the Lorentz curvature has components
\be
\begin{array}{lcl}
R^{\td\,\tc}_{\d\,\g\, ba}\ =\
2\eps_{\d\g}\textrm{tr}(\s_{dc}\s_{ba})F^{dc[\td\tc]}e^{-\f}&&
R_{\td\,\tc}^{\dd\,\dg}{}_{ ba}\ =\
2\eps^{\dd\dg}\textrm{tr}(\bs_{dc}\bs_{ba})F^{dc}{}_{[\td\tc]}e^{-\f}\\[2mm]
R_{\td\,\g\, ba}^{\dd\,\tc}\ =\
-4\eps_{dcba}U^d{}^\td{}_\tc(\s^c\e)_\g{}^\dd&& R_{\cc\au\,ba}\ =\
0\\[2mm]
R^\td_\d{}_c{}_{ba}\ =\
-2i(\s_c)_{\d\da}(\eps\bs_{ba})_{\dg\db}\S^{(\dg\db\da)\td}
&&R_\td^\dd{}_c{}_{ba}\ =\
-2i(\bs_c)^{\a\dd}(\eps\s_{ba})^{\g\b}\S_{(\g\b\a)\td}
\\[2mm]
\quad-\frac{i}{2}\textrm{tr}(\s_{ba}\s_{ef})(\s_c\bla_\ta)_\d
F^{ef}{}^{[\td\ta]}e^{-\f}
&&\quad-\frac{i}{2}\textrm{tr}(\bs_{ba}\bs_{ef})(\bs_c\la^\ta)^\dd
F^{ef}{}_{[\td\ta]}e^{-\f}
\\[2mm]
\quad+\frac{i}{4}(\s_c\bs_e\s_{ba}\la^\td)_\d P^e
&&\quad+\frac{i}{4}(\bs_c\s_e\bs_{ba}\bla_\td)^\dd \bP^e
\end{array}
\ee
and
\bea
R_{dc}{}_{ba}&=&(\eps\s_{dc})^{\d\g}(\eps\s_{ba})^{\b\a}V_{(\d\g\b\a)}
+(\eps\bs_{dc})_{\dd\dg}(\eps\bs_{ba})_{\db\da}V^{(\dd\dg\db\da)}\nn\\[2mm]
&&+\frac{1}{2}\left(\eta_{db}R_{ca}-\eta_{da}R_{cb}+\eta_{ca}R_{db}-\eta_{cb}R_{da}\right)
-\frac{1}{6}(\eta_{db}\eta_{ca}-\eta_{da}\eta_{cb})R
\eea
with the supercovariant Ricci tensor, $R_{db}=R_{dcba}\eta^{ca}$, given by
\bea
R_{db}\  &=& -2\cd_d\f\,\cd_b\f-\frac{1}{2}e^{-4\f}H^*_d\,H^*_b
-e^{-2\f}F_{df[\tb\ta]}F_b{}^f{}^{[\tb\ta]}+\frac{1}{4}\eta_{db}e^{-2\f}F_{ef[\tb\ta]}F^{ef[\tb\ta]}\nn\\[2mm]
&&+\frac{1}{8}\sum_{db}\left\{i(\cd_b\la^\tf)\s_d\bla_\tf-i\la^\tf\s_d\cd_b\bla_\tf
+e^{-2\f}H_d^*(\la^\tf\s_b\bla_\tf)\right\}\nn\\[2mm]
&&-\frac{1}{32}(\la^\tf\s_d\bla_\tf)\,(\la^\ta\s_b\bla_\ta)-\frac{1}{16}\eta_{db}(\la^\ta\la^\tf)(\bla_\ta\bla_\tf)
\label{Riccitensor}
\eea
and the corresponding Ricci scalar, $R=R_{db}\eta^{db}$, which is then
\be
R=-2\cd^a\f\cd_a\f-\frac{1}{2}H^*{}^aH^*{}_ae^{-4\f}
+\frac{3}{4}e^{-2\f}H^*{}^a(\la^\ta\s_a\bla_\ta)
+\frac{3}{8}(\la^\tb\la^\ta)(\bla_\tb\bla_\ta).
\ee
The tensors $V_{(\d\g\b\a)}$ and $V^{(\dd\dg\db\da)}$ are components
of the usual Weyl tensor. Like the
gravitino Weyl tensors, $\S_{(\g\b\a)\ta}$ and $\S^{(\dg\db\da)\ta}$,
their lowest components do not
participate in the equations of motion.

As for the 2--form sector, besides the supercovariant field strength
of the antisymmetric tensor,
$H_{cba}$, the non-zero components of the 3--form $H$, which do not
have central charge indices, are
\be
H^\dg_\tc{}^\tb_\b{}_a=-2i\d^\tb_\tc(\s_a\eps)_\b{}^\dg e^{2\f}
\qquad H^\tc_\g{}_{ba}\ =\ 4(\s_{ba}\la^\tc)_\g e^{2\f}
\qquad H_\tc^\dg{}_{ba}\ =\ 4(\bs_{ba}\bla_\tc)^\dg e^{2\f}\ .
\ee
The components with at least one central charge index, are related to
the torsion components by
\be
H_{\cd\cc\au}\ =\ T_{\cd\cc}{}^\az\gg_{\az\au},
\ee
with the metric $\gg_{\az\au}$ defined in \equ{metric}.

\end{appendix}

\bibliography{newREF}
\bibliographystyle{JHEP}
\end{document}